\documentclass[conference]{IEEEtran}
\IEEEoverridecommandlockouts
\ifCLASSINFOpdf
  \usepackage[pdftex]{graphicx}
  \graphicspath{{./images/}}
\else
  \usepackage[dvips]{graphicx}
  \graphicspath{{./images/}}
\fi

\usepackage{amsmath}

\usepackage{algorithmic}

\ifCLASSOPTIONcompsoc
  \usepackage[caption=false,font=normalsize,labelfont=sf,textfont=sf]{subfig}
\else
  \usepackage[caption=false,font=footnotesize]{subfig}
\fi

\begin{document}
%
\title{A Practical Approach for Rate-Distortion-Perception Analysis in Learned Image Compression}

\author{\IEEEauthorblockN{Ogun Kirmemis, A. Murat Tekalp$^*$\thanks{$^*$This work was supported by TUBITAK 1001 project 217E033, TUBITAK 2247-A National Leader Researchers Award No. 120C156, and Turkish Academy of Sciences (TUBA).} }
\IEEEauthorblockA{Electrical and Electronics Engineering\\
Koc University, Istanbul, Turkey\\
Email: okirmemis16@ku.edu.tr, mtekalp@ku.edu.tr}
}


\maketitle

\begin{abstract}
Rate-distortion optimization (RDO) of codecs, where distortion is quantified by the mean-square error, has been a standard practice in image/video compression over the years. RDO serves well for optimization of codec performance for evaluation of the results in terms of PSNR. However, it is well known that the PSNR does not correlate well with perceptual evaluation of images; hence, RDO is not well suited for perceptual optimization of codecs. Recently, rate-distortion-perception trade-off has been formalized by taking the Kullback–Leibler (KL) divergence between the distributions of the original and reconstructed images as a perception measure. Learned image compression methods that simultaneously optimize rate, mean-square loss, VGG loss, and an adversarial loss were proposed. Yet, there exists no easy approach to fix the rate, distortion or perception at a desired level in a practical learned image compression solution to perform an analysis of the trade-off between rate, distortion and perception measures. In this paper, we propose a practical approach to fix the rate to carry out perception-distortion analysis at a fixed rate in order to perform perceptual evaluation of image compression results in a principled manner. Experimental results provide several insights for practical rate-distortion-perception analysis in learned image compression.
\end{abstract}
	


%
\IEEEpeerreviewmaketitle

\section{Introduction}
The third deep learning revolution, which started in the early 2010s, not only changed the course of artificial intelligence and machine learning but also had a serious impact on the signal, image, and video processing. One of the first successful applications of deep learning to image processing was in single-image super-resolution (SISR) \cite{srcnn}, which started the new era of learned SISR techniques.  
Deep learning also had a significant impact on image/video compression enabling end-to-end optimization of image/video codecs. Early works on end-to-end optimized learned codecs considered traditional rate-distortion optimization, while more recent works have started to investigate rate-distortion-perception optimization.

Initial works on deep learned image compression relied on pixel-recurrent RNNs, where Toderici~\textit{et al.}~\cite{toderici} managed to exceed the performance of JPEG. Later, it became clear that better performance can be achieved by end-to-end rate-distortion optimized learned image compression based on jointly learning the parameters of a variational autoencoder~\cite{balle2017endtoend} and an entropy model. The~state-of-the-art in learned image compression has advanced fast after 2018 through annually organized Challenge on Learned Image Compression (CLIC), where the participants were asked to compress images below 0.15 bpp and have better performance than BPG in terms of either peak signal-to-noise ratio (PSNR) or structural similarity index measure (SSIM). There are several advantages of end-to-end optimized learned image compression compared to traditional engineered codecs. First, we are not restricted to using linear and block-based transforms and engineered context models for arithmetic coding. Second, the whole system can be optimized end-to-end according to a desired non-convex (but differentiable) loss function, which can be any combination of $l1/l2$ loss, a feature-driven visual distortion loss, such as SSIM, or a no-reference perceptual loss.

Compression algorithms are optimized and evaluated based on Shannon's rate-distortion framework, where the goal is to minimize the distortion, typically represented by the mean-square error (MSE), at a given bitrate or vice versa. Although this framework helps design codecs that yield the best PSNR performance at a given bitrate, it is well known that PSNR does not correlate well with human perception of image quality since optimizing MSE or PSNR can result in blurry images. The standard approach for perceptual quality evaluation in the image compression community is to conduct subjective tests based on the Mean Opinion Score (MOS). However, this is a time-consuming process, it is not scalable to large datasets, and the results have not been optimized for perceptual quality. 
With the advent of learned image compression, we can now end-to-end optimize codecs with respect to perceptual distortion losses, such as SSIM~\cite{ssim}, or no-reference perceptual losses in addition to MSE. Recent works of Matsumoto \cite{Matsumoto_2018,Matsumoto_2019} and Blau \textit{et al.}~\cite{rdp} have shown that there exist a triple trade-off between rate, distortion (fidelity), and perception (RDP). However, there exist two important challenges with employing rate-distortion-perception optimization in practice: i) there is no universally agreed objective measure of perceptual quality, ii) there is no practical approach to fix or control rate, distortion, and perception separately in order to employ this framework in learned codecs at the moment.

In this paper, we introduce a practical approach to perform rate-distortion-perception analysis in learned image compression. In particular, we perform perception-distortion analysis at various fixed bitrates sequentially. To this effect, we fix the bitrate by freezing the training of the encoder and hyperprior networks at successive desired bitrate points. Once the best perception-distortion point is found at a fixed rate, we continue training the encoder and hyperprior networks to reach another rate point. We continue this procedure of finding the best perception-distortion points at certain fixed bitrates until the set of desired rate points are covered.
The remainder of the paper is organized as follows: Section~\ref{sec:related} discusses related works. The details of the proposed procedure are presented in Section~\ref{method}. Experimental results are shown in Section~\ref{results}. Finally, Section~\ref{conc} concludes the paper.


\section{Related Work}
\label{sec:related}
Ballé \textit{et al.}~\cite{balle2017endtoend, balle2018scale,balle2021} laid the foundation of end-to-end optimized image compression using an auto-encoder by introducing joint optimization of rate, distortion, and a probability model. They estimate the rate by the entropy of a non-parametric prior distribution on the latent representation. During the training, they model quantization by adding a uniform distributed noise to deal with the non-differentiability of quantization. Ballé~\textit{et al.}~\cite{balle2018scale} introduced forward adaptation of the entropy model by a learnable scale hyperprior, where they assume the latents are independent Laplace random variables and estimate the scale of the variables through a hyperprior network. Minnen~\textit{et al.}~\cite{minnen2018} included a backward adaptation step by adding a context model, where they used PixelCNN~\cite{pixelcnn} architecture for pixel-wise context model. In addition to the scale, their network also estimated the mean of latent representation. However, PixelCNN dramatically slows the decoding process, because it gives a single pixel output for a channel at a time. This problem is addressed by the channel-wise autoregressive entropy model~\cite{minnen2020channelwise}, which slices the latent representation in the channel dimension and every slice is conditioned on the prior slice. This process is easily parallelizable due to the nature of CNNs; hence, it is much faster than the PixelCNN. These works have considered optimization with respect to MSE or SSIM losses; hence, they did not explicitly address the perception dimension.

Although the works of Matsumoto~\cite{Matsumoto_2018,Matsumoto_2019} and Blau \textit{et al.}~\cite{rdp} have introduced the rate-distortion-perception trade-off framework, they did not propose practical approaches to achieve combined optimization of rate, distortion, and perceptual measures. 
Mentzer \textit{et al.}~\cite{hific} proposed optimization of a combination of rate, MSE, and perceptual distortion measure, together with an adversarial loss in order to encode images at very low bitrates with an acceptable rate-distortion-perception trade-off. However, they employed ad-hoc tricks to keep the rate approximately constant.
Prior works on generative image compression include \cite{rippel2017realtime} and \cite{santurkar2017generative}. One advantage of generative compression is that it can encode images with extremely low bitrates (less than 0.1 bpp) and decode perceptually meaningful images since GANs are able to learn the image manifold and create real looking images~\cite{gan}. Agustsson \textit{et al.}~\cite{agustsson2019generative} propose using conditional adversarial training which helps decoder to identify which regions to hallucinate and which regions to decode faithfully. However, these methods did not consider a formal approach for rate-distortion-perception trade-off and decoded images might not be faithful to the original images by some desired amount. Our main contribution in this paper is to propose a practical approach to fix the rate exactly at the desired value for a proper rate-distortion-perception analysis.


\section{Rate-constrained optimization of Perception and Distortion}
\label{method}
A traditional codec can be optimized for a hard constrained rate-distortion optimization with an exhaustive search of compression parameters. However, their neural network counterparts are trained with soft constrained loss functions. This leads to the same model achieving different rate-distortion points on different images. Ballé \textit{et al.} optimizes the rate-distortion trade-off by minimizing the following equation:
\begin{equation}
    L =  d(x,\hat{x}) + \lambda H(y)
    \label{eq:loss_rd}
\end{equation}
where $x$ is the original image, $\hat{x}$ is the decoded image, $d(x,\hat{x})$ denotes a distortion function, such as the MSE, and $H(y)$ is the entropy (bitrate) of the compressed representation $y$.

Inclusion of a perceptual loss term to the overall loss function further complicates the optimization process since we need to weigh the three terms with proper weights. There are different approaches to optimizing these quantities. 

Approach taken by Agustsson~\textit{et al.}~\cite{agustsson2019generative} is assuming an upper bound on the entropy of the latent code. When the number of quantization levels is finite and the dimension of the latent code is fixed, the upper bound on the entropy can be estimated by assuming the latent code has independent uniform distributions. It should be noted that this would be a very loose bound since the actual latent code would be neither independently nor uniformly distributed. Perceptual loss is added in the means of adversarial loss where a discriminator $D$ is added to differentiate between the decoded image and the original image. Due to adversarial training, two different  loss functions has to be optimized concurrently. These loss functions are:
\begin{equation}
    L =  d(x,\hat{x}) + \lambda H(y) - \beta log(D(\hat{x}, y))
    \label{eq:loss_gen}
\end{equation}
\begin{equation}
    L_D = -log(1-D(\hat{x}, y)) - log(D(x,y))
    \label{eq:loss_disc}
\end{equation}
The new terms in (\ref{eq:loss_gen}) come from the discriminator $D$, $\lambda$ and $\beta$ controls the trade-off between rate, distortion and perception. Equation (\ref{eq:loss_disc}) is the standard binary cross-entropy loss used for training the discriminator.

Agustsson~\textit{et al.}~\cite{hific} also uses the loss functions (\ref{eq:loss_gen}) and (\ref{eq:loss_disc}); however, with a significant difference: In \cite{agustsson2019generative} $\lambda$ is set to 0 and thus the rate is not optimized directly. Instead, as mentioned above the authors rely on the loose upper bound. Since the entropy is not directly optimized, a distribution on the latent code cannot be forced as in the RDO framework provided by~\cite{balle2017endtoend}. This approach simplifies optimization greatly. On the other hand, \cite{hific} uses both loss functions in (\ref{eq:loss_gen}) and (\ref{eq:loss_disc}) with non-zero hyperparameters. However, with this formulation, optimizing for different bitrates is more difficult since changing one of the hyperparameters in (\ref{eq:loss_gen}) actually changes the relative weight of the other two terms. In order to constrain the rate, authors use two hyperparameters $\lambda ^{a}$ and $\lambda ^{b}$ with $\lambda ^{a} \gg  \lambda ^{b}$ to change the weight of the entropy term adaptively. If the target rate is exceeded then $\lambda ^{a}$ is used to bring the model back to the target bitrate, otherwise $\lambda^{b}$ is used.

Our approach is different from the aforementioned methods since we fix the bitrate by freezing the encoder. First, we train the network with (\ref{eq:loss_rd}) to achieve a target bitrate. Since the entropy only depends on the encoder and the entropy model defined by the hyperprior network, the bitrate can be fixed by freezing these two networks. At this point, we can optimize for distortion-perception trade-off directly since we have a single hyperparameter in the loss function:
\begin{equation}
    L= \gamma d(x,\hat{x}) + L_p(\hat{x}, x)
    \label{eq:loss_pd}
\end{equation}
where $L_p$ is a generic perceptual loss, $\gamma$ is a hyperparameter controlling the perception-distortion trade-off.

The proposed procedure for Distortion-Perception optimization at a fixed bitrate with our practical approach for fixing the bitrate can be summarized as:
\begin{algorithmic}[1]
\STATE Randomly initialize the network
\STATE Train the network to reach the desired rate-distortion point using (\ref{eq:loss_rd})
\STATE Freeze the weights of the encoder and entropy model estimator sub-networks to fix the encoding rate
\STATE Start the search for optimal $\gamma$ with the loss function in (\ref{eq:loss_pd})
\STATE Find the $\gamma$ that corresponds to the knee point (see Figures \ref{fig:lpipsvspsnr} and \ref{fig:lpipsvsssim})
\end{algorithmic}

This procedure is repeated for as many different rate points as desired.

\section{Experimental Results}
\label{results}
In this section, we provide experimental results to demonstrate that there exists a best perception-distortion trade-off point when encoding at a fixed bitrate. To this effect, we generate multiple decoded images from the same encoded bitstream at a fixed rate by optimizing the decoder network parameters according to different weighting of distortion and perceptual loss functions according to the procedure described above.

B{\'e}gaint \textit{et al.} \cite{compressai} provides pretrained models of \cite{minnen2018} for eight different bitrates. We picked the models with the four lowest bitrates since the perception-distortion trade-off is more apparent in the low bitrate regime as suggested by \cite{rdp}. The encoder and the entropy models are frozen before training. For the perceptual loss, we employed LPIPS v0.1~\cite{lpips} with VGG features~\cite{vgg}. VGG features are used widely for perceptual improvement since Ledig~\textit{et al.}~\cite{srgan} showed that the use of VGG loss improves MOS significantly. However, perceptual evaluation is done with LPIPS with AlexNet features\cite{alexnet} since authors state that this network correlates more with MOS. Using a different perceptual metric in test time also provides some impartialness in evaluation. The network is trained with Adam optimizer \cite{adam} with a learning rate of $10^{-5}$ and default parameters. We train the network for 400000 iterations on Vimeo-90k dataset~\cite{vimeo90k} with a batch size of 16 on $256\times 256$ patches.

\begin{figure}[!ht]
\centering
\includegraphics[width=0.55\textwidth]{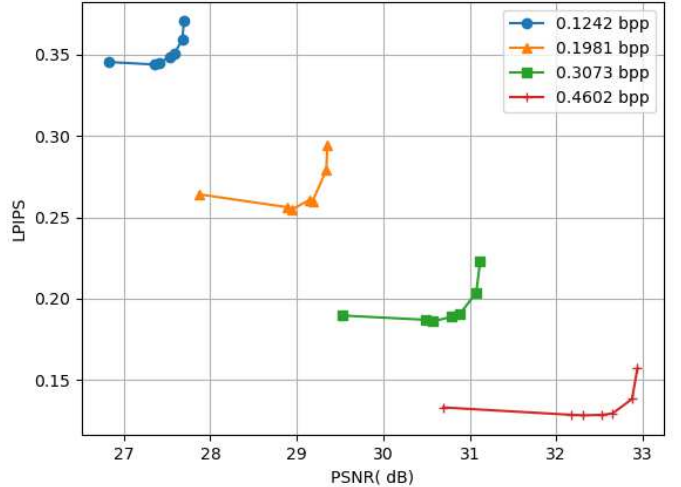}
\caption{LPIPS vs. PSNR plots of models trained at different fixed bitrates on Kodak dataset. The~best perception (LPIPS)-distortion (PSNR) trade-off point at each fixed bitrate are determined as the knee-point of the respective curves.}
\label{fig:lpipsvspsnr}
\end{figure}

\begin{figure}[!ht]
\centering
\includegraphics[width=0.55\textwidth]{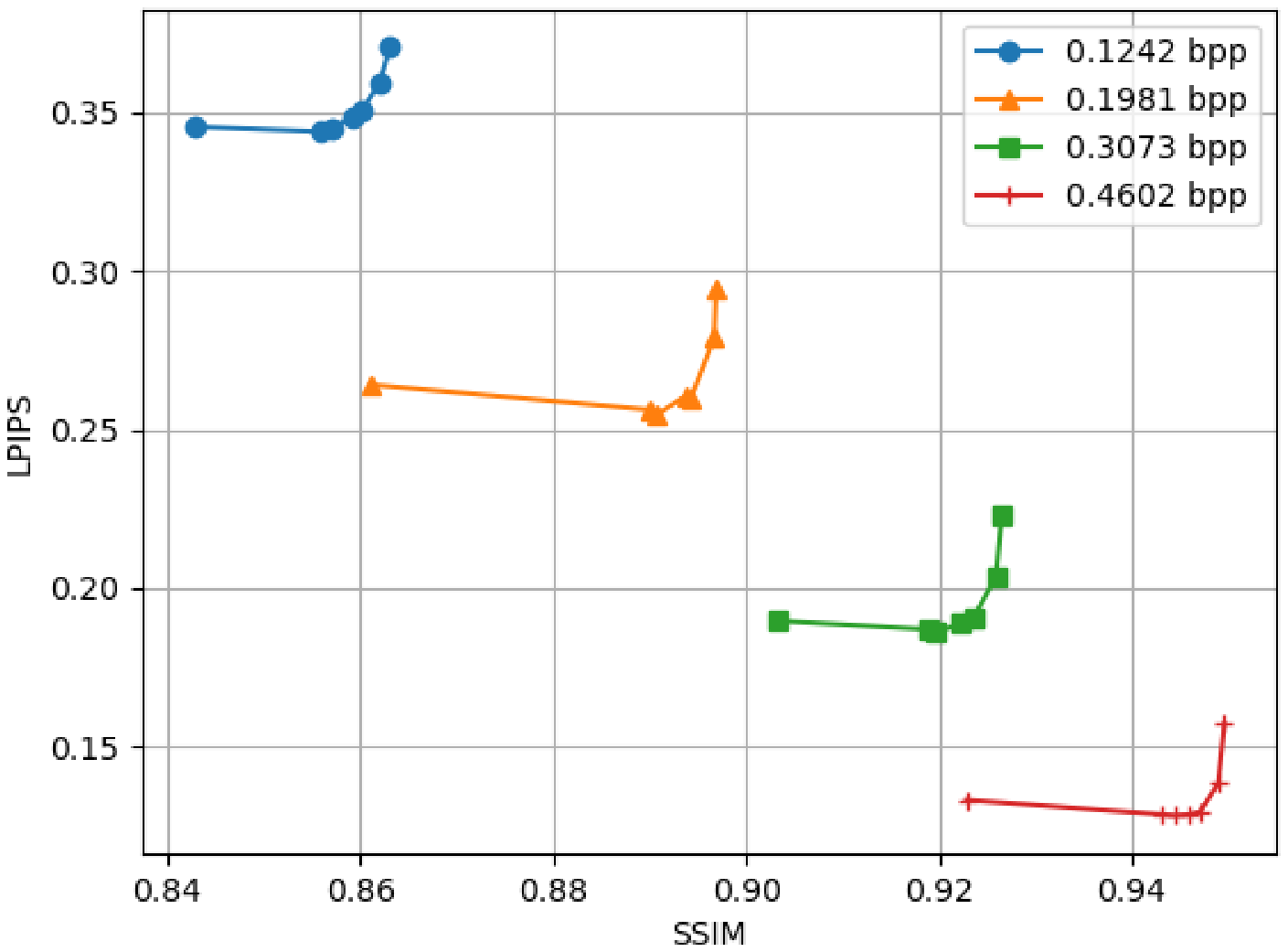}
\caption{LPIPS vs. SSIM plots of models trained at different fixed bitrates on Kodak dataset. The~best perception (LPIPS)-distortion (SSIM) trade-off point at each fixed bitrate are determined as the knee-point of the respective curves.}
\label{fig:lpipsvsssim}
\end{figure}

\begin{figure*}[ht]
\centering
\subfloat[]{\includegraphics[width=6cm]{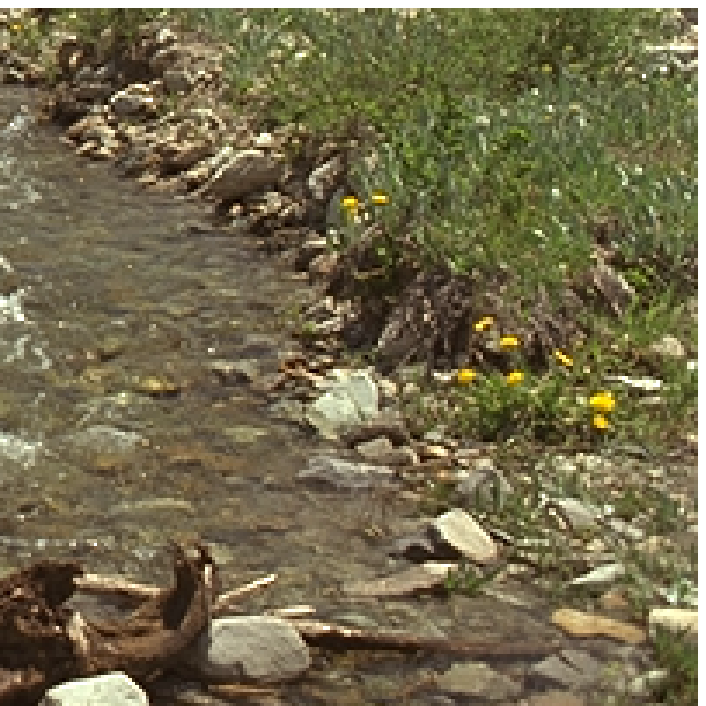}\label{fig:orig13}}\hfill
\subfloat[]{\includegraphics[width=6cm]{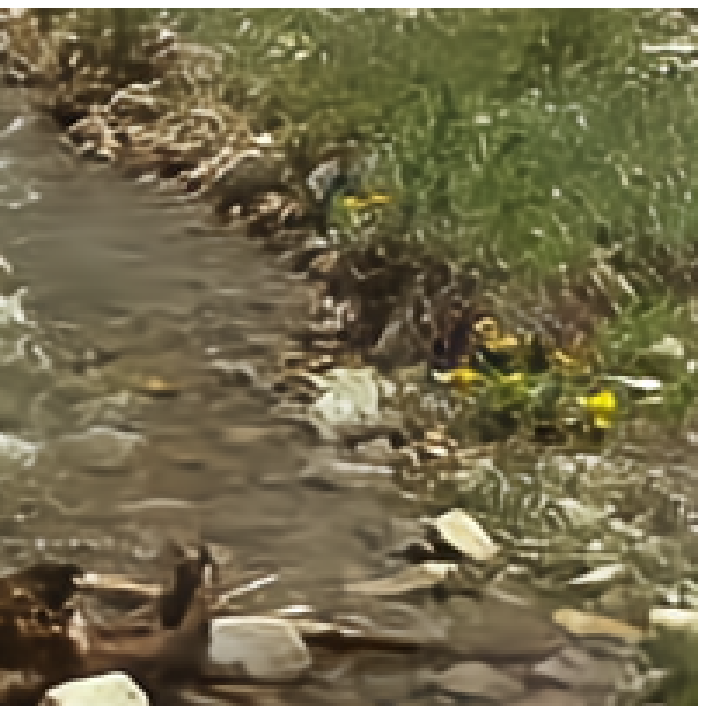}\label{fig:kodak13pre}} \hfill
\subfloat[]{\includegraphics[width=6cm]{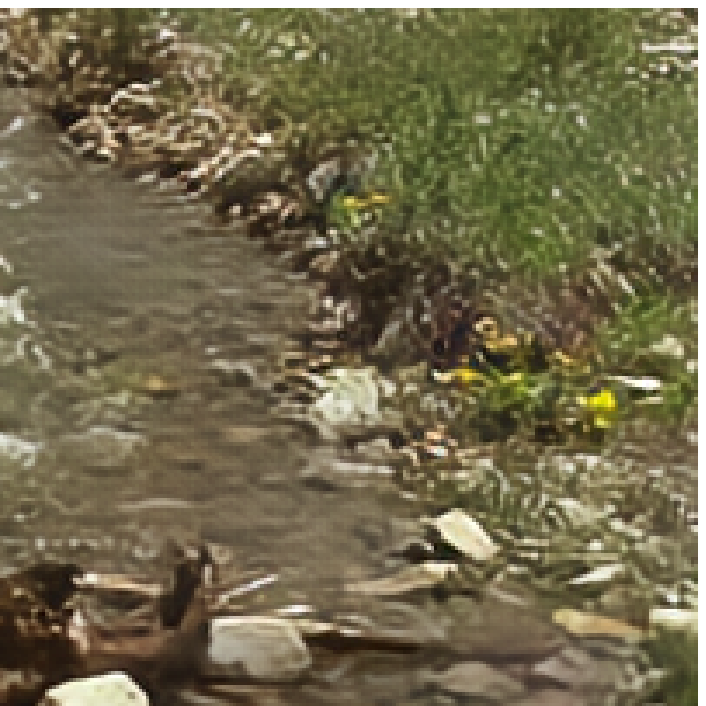}\label{fig:kodak13p}}
\hfill
\caption{Example crop from Kodak dataset: (a) original image, (b) output of decoder model optimized with respect to MSE only, (c) output of decoder model with $\gamma=5\times 10^{-4}$. Both images (b) and (c) are reconstructed from the same encoded bitstream at 0.40 bpp. PSNR of (b) and (c) are 24.58 dB and 24.35 dB, respectively. SSIM scores are also similar: 0.7967 and 0.7920, respectively. Although (b) has slightly less distortion in terms of MSE, (c) looks sharper especially on the grass and the stones in the water. This is evident from the LPIPS scores: (b) scored 0.41 whereas (c) scored 0.35. (Lower LPIPS is better.)}
\label{fig:kodak13res}
\end{figure*}

Figure \ref{fig:lpipsvspsnr} shows perception-distortion trade-off in terms of LPIPS and PSNR. Different curves represent the different bitrates. For every curve, the top right point belongs to the pretrained model which is trained only with (\ref{eq:loss_rd}). Since the pretrained model is not optimized for perception, we call this network MSE only. After freezing the encoder side and hyperprior sub-networks, we start the perception-distortion optimization. The other points on the curves are the results of the experiments with $\gamma$ values of $10^{-2}$, $10^{-3}$, $5\times 10^{-4}$, $10^{-4}$, $5\times10^{-5}$, and $0$ from right to left, respectively. $\gamma=10^{-3}$ improves LPIPS at the cost of a slight decrease in PSNR. The model trained with $\gamma=10^{-4}$ still shows some improvement in LPIPS but marginal gain is much lower, also evident from the slope between the points $\gamma=10^{-3}$ and $\gamma=10^{-4}$. The extreme case of $\gamma=0$ is trained to display the nature of the optimization problem. Since MSE minimizes the pixel-wise differences, the network's outputs start to differ too much from the input without MSE loss, this becomes harmful for both PSNR and LPIPS. Figure \ref{fig:lpipsvspsnr} shows the optimal point to stop the hyperparameter search is at the knee point of the curves.

Figure \ref{fig:lpipsvsssim} shows the perception-distortion trade-off in terms of LPIPS and SSIM. Although we have not explicitly optimized for SSIM, our observations on Figure \ref{fig:lpipsvspsnr} also applies here. After reaching the knee of the curve, decreasing $\gamma$ more harms the network since it only increases distortion without an improvement in perception. Both figures clearly show there is an optimal perception-distortion point for every fixed bitrate. 

\begin{figure*}[ht]
\centering
\subfloat[]{\includegraphics[width=6cm]{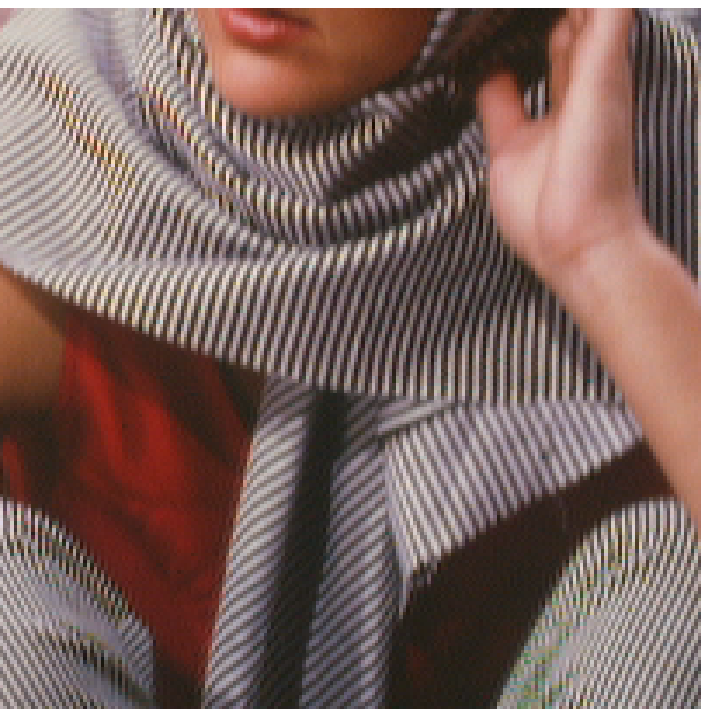}\label{fig:orig02}}\hfill
\subfloat[]{\includegraphics[width=6cm]{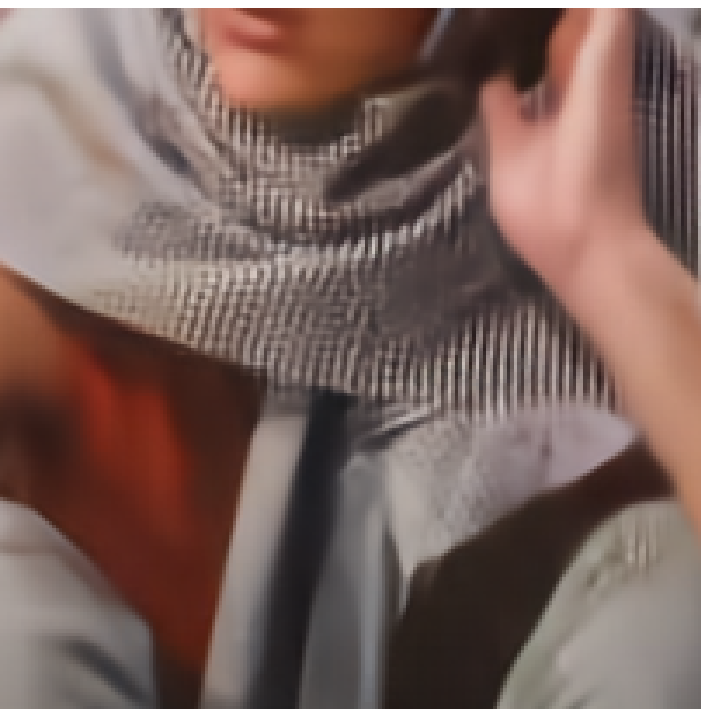}\label{fig:set02pre}} \hfill
\subfloat[]{\includegraphics[width=6cm]{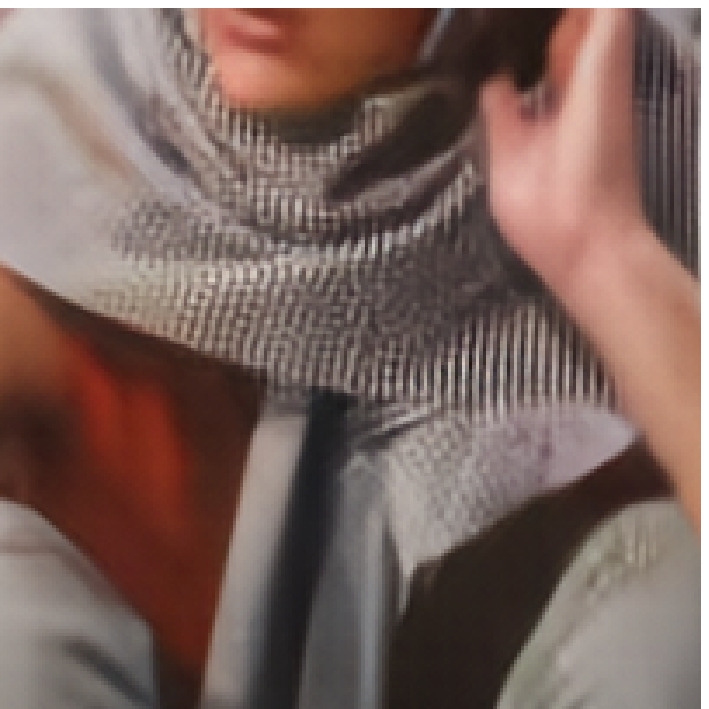}\label{fig:set02p}}
\hfill
\caption{Example crop from Set14 dataset~\cite{Dataset:set14}: (a) original image, (b) output of decoder model optimized with respect to MSE only, (c) output of decoder model with $\gamma=5\times 10^{-4}$. Both images (b) and (c) are reconstructed from the same encoded bitstream at 0.16 bpp. PSNR of (b) and (c) are 25.14 dB and 25.11 dB, respectively. SSIM scores are 0.8253 and 0.8243, respectively. Although (b) has slightly less distortion in terms of MSE, (c) looks sharper in the area on the left side of the woman's hand. This is also evident from the LPIPS scores: (b) scored 0.359 whereas (c) scored 0.347. (Lower LPIPS is better.)}
\label{fig:set02res}
\end{figure*}

Figure \ref{fig:kodak13res} shows an example image from Kodak dataset compressed with the trained networks. The compressed images are at the same bitrate and have similar distortion levels, however, Figure \ref{fig:kodak13p} looks sharper on the grass and rocks despite having higher distortion in terms of both PSNR and SSIM. These experimental results clearly show that our optimization method can optimize for rate-distortion-perception. Figure \ref{fig:set02res} shows another example from Set14 dataset~\cite{Dataset:set14}. The stripes on the scarf of the woman are lost on both compressed images on the left side, however, MSE trained network also blurred the region down the woman's palm whereas the perceptually trained network looks sharper. Figures \ref{fig:kodak13res} and \ref{fig:set02res} shows that our approach works for different bitrates.

\section{Conclusion}
\label{conc}
This paper has two main contributions: 
\begin{itemize}
\item We propose a practical approach to fix the bitrate of the encoded image at a desired value to perform perception-distortion analysis at a fixed bitrate.
\item We propose a principled approach to determine the best perception-distortion trade-off point at a fixed rate. 
\end{itemize}
Experimental results demonstrate that there exists a best perception-distortion trade-off operating point for each fixed bitrate.
We also propose a framework and a procedure to conduct perception-distortion analysis at a succession of fixed bitrates to achieve the best rate-distortion-perception optimization for learned image compression. In the future, we would like to extend our work to include GAN based networks.
\vspace{12pt}






{
\bibliographystyle{IEEEtran}
\bibliography{ieeeabrv,egbib}
}
\end{document}